\documentclass[runningheads]{cl2emult}

\usepackage{makeidx}  
\usepackage{graphicx} 
\usepackage{subeqnar} 
\usepackage{multicol} 
\usepackage{cropmark} 
\usepackage{eso}      
\makeindex            

\def\ltsima{$\; \buildrel < \over \sim \;$}
\def\simlt{\lower.5ex\hbox{\ltsima}}            
\def\gtsima{$\; \buildrel > \over \sim \;$}
\def\simgt{\lower.5ex\hbox{\gtsima}}            

\begin{document}
\title*{Mining the Blazar Sky}
\toctitle{Mining the Blazar Sky}
%
%
\titlerunning{Mining the Blazar Sky}
%
\author{Paolo Padovani\inst{1,2,3}
\and Paolo Giommi\inst{4}}
\authorrunning{Paolo Padovani \& Paolo Giommi}
%
%
\institute{Space Telescope Science Institute, 3700 San Martin Drive, Baltimore,
MD, 21218, USA
\and Affiliated to the Astrophysics Division, Space Science Department, 
European Space Agency
\and On leave from Dipartimento di Fisica, II Universit\`a di Roma ``Tor 
Vergata'', \\
Via della Ricerca Scientifica 1, I-00133 Roma, Italy
\and BeppoSAX Science Data Center, ASI, Via Corcolle 19, I-00131 Roma, Italy}

\maketitle              

\begin{abstract}
We present \index{abstract} the results of our methods to ``mine'' the blazar
sky, i.e., select blazar candidates with very high efficiency. These are based
on the cross-correlation between public radio and X-ray catalogs and have 
resulted in two surveys, the Deep X-ray Radio Blazar Survey (DXRBS) and the
``Sedentary'' BL Lac survey. We show that data mining is vital to select
sizeable, deep samples of these rare active galactic nuclei and we touch upon
the identification problems which deeper surveys will face.
\end{abstract}

\section{The Importance of Being a Blazar}

The current paradigm for Active Galactic Nuclei (AGN) \index{AGN} includes a
central engine, possibly a massive black hole, surrounded by an accretion disk
and by fast-moving clouds, probably under the influence of the strong
gravitational field, emitting Doppler-broadened lines. More distant clouds
emit narrower lines. Absorbing material in some flattened configuration
(usually idealized as a toroidal shape) obscures the central parts so that for
transverse lines of sight only the narrow-line emitting clouds are seen. In
radio-loud objects we have the additional presence of a relativistic jet,
roughly perpendicular to the disk. This produces strong anisotropy and
amplification of the continuum emission (``relativistic beaming'') when viewed
face-on. Within this scheme, blazars represent the fraction of AGN with their
jets at relatively small ($\simlt 20 - 30^{\circ}$) angles w.r.t. the line of
sight (e.g., \cite{up95}).

Given that extragalactic jets are relatively narrow, it is relatively unlikely
that our line of sight will intercept a jet. This, together with the fact that
radio-loud sources constitute only $\sim 10\%$ of AGN, implies that blazars
represent a rare class of objects, making up considerably less than 5\% of all
AGN \cite{p97}. \index{blazars}

The blazar class includes flat-spectrum radio quasars (FSRQ) and BL Lacertae
objects. Within the so-called ``unified schemes'', these are thought to be the
``beamed'' counterparts of high- and low-luminosity radio galaxies,
respectively. The main difference between the two blazar classes lies in their
emission lines, which are strong and quasar-like for FSRQ and weak or in most
cases absent in BL Lacs. 

In addition to their rareness, blazars are the most extreme variety of AGN
known. Their main properties include: 1. smooth, broad-band, non-thermal
continuum, covering the whole electromagnetic spectrum (radio to
$\gamma$-rays); 2. compact (core flux $\gg$ extended flux), flat-spectrum
(radio spectral index $\alpha_{\rm r} \simlt 0.5$), radio morphology; 3. rapid
variability (large $\Delta L/\Delta t$); 4. high and variable optical
polarization; 5. superluminal motion in sources with
multiple-epoch Very Large Baseline Interferometry (VLBI) maps. 


The last property might require some explanation. The term ``superluminal
motion'' describes proper motion of source structure (traditionally mapped at
radio wavelengths) that, when converted to an apparent speed $v_{\rm app}$,
gives $v_{\rm app} > c$. This phenomenon occurs for emitting regions moving at
very high (but still $< c$) speeds at small angles to the line of
sight \cite{r66}.

In a nutshell, blazars are sites of very high energy phenomena, both in terms
of photon energies, reaching the TeV ($\sim 2 \times 10^{26}$ Hz) range, and
bulk motion, with Lorentz factors ($\Gamma = (1-\beta^2)^{-1/2}$, with $\beta
= v/c$) up to $\Gamma \sim 40$ (or speeds of the emitting material reaching
0.9997 the speed of light).

The broad, strong, continuum of blazars is very relevant to data
mining, as blazars will show up in every catalog at all wavelengths.
Moreover, their rareness implies that data mining is vital to assemble
relatively large samples. On both accounts, blazars represent ideal test cases
for data mining studies.

\section{Finding Blazars}

As blazars are rare, ``pencil beam'' surveys are not suited to find them;
large areas are needed. Moreover, BL Lac spectra are almost featureless, so
these sources are also hard to identify. As a consequence, all existing blazar
samples were, until recently, relatively small and at high fluxes.

The small sample size means that the derivation of the beaming parameters
(Lorentz factors, angles w.r.t. the line of sight) based on luminosity
function studies (e.g., \cite{pu90}, \cite{ups91}) is considerably uncertain,
especially at low powers. The high fluxes imply that we do not know if the
relativistic beaming scenario, which appears to work reasonably well for the
available samples, still applies at lower fluxes and powers. In other words,
our understanding of the blazar phenomenon is mostly based on a relatively
small number of intrinsically luminous sources, which means we have only
sampled the tip of the iceberg of the blazar population.

The need for deeper, larger blazar samples is then obvious. But how to fulfill
that need? The ``classical'' approach, i.e., obtain an optical spectrum of all
sources to identify them, can be applied to large-area, shallow surveys, as
these include a manageable number of objects (say up to a thousand or
so\footnote{Dedicated instruments or projects, like the Two degree Field (2dF;
\cite{sh}, \cite{dp}) and the Sloan Digital Sky Survey (SDSS; \cite{th}), can
actually adopt the classical approach for a much larger number of sources (of
the order of 250,000 for 2dF and a million for SDSS). This, however, requires
populations with relatively large surface density (2dF) and large investments
(SDSS). In both cases the optical limit is relatively high ($\sim$20--21
magnitude).}). This is an obviously long process but it can be completed in a
reasonable amount of time. For example, in the case of the {\it Einstein}
Medium Sensitivity Survey (EMSS), which includes 835 sources, all candidates
in the large X-ray error boxes had to be observed to identify the most likely
X-ray source. This process took about 10 years to complete. \index{NVSS}

When dealing with deeper surveys, one runs into problems. This is vividly
illustrated by comparing the 1 Jy catalog \cite{k81}, which covers 
the whole sky off the Galactic plane at 5GHz and the NRAO VLA Sky Survey
(NVSS) \cite{c98}, which covers the sky north of $\delta = -40^{\circ}$ at 1.4
GHz. The area covered by the two surveys is more or less the same but the
latter goes almost three orders of magnitude deeper in flux. As a result the
total number of sources increases by a factor $\sim 3,500$, going from 527 to
almost 2 million. It is clear that this requires a radical change in the way
source identification is carried out.


Imagine in fact to look for blazars in the NVSS survey. Identifying all of the
1.8 million sources would be impossible on a reasonable timescale,
even with unlimited access to telescope time (note that $\sim 10\%$ of the 1
Jy sources are still unidentified). Hence the need to increase the selection
efficiency to restrict the number of blazar candidates down to a manageable
size, allowing at the same time the selection of a well-defined sample
suitable for statistical analysis. And this is where data mining comes into
play.

We present here two ``real-life'' applications of data mining to the selection
of blazars, to assemble deeper samples and to select ``extreme''
sources based on their location in parameter space. 

\section{The Deep X-ray Radio Blazar Survey (DXRBS)}

The basic idea behind the Deep X-ray Radio Blazar Survey (DXRBS) is simple.
Blazars are relatively strong X-ray and radio emitters so selecting X-ray and
radio sources with flat radio spectrum (one of their defining properties)
should be a very efficient way to find these rare sources. By adopting a
spectral index cut $\alpha_{\rm r} \le 0.7$ DXRBS: 1. selects all FSRQ
(defined by $\alpha_{\rm r} \le 0.5$); 2. selects basically 100\% of BL Lacs;
3. excludes the large majority of radio galaxies. DXRBS uses a
cross-correlation of all serendipitous X-ray sources in the publicly available
{\it ROSAT} database WGACAT \cite{wga}, with a number of publicly available
radio catalogs (GB6, NORTH20, PMN). \index{DXRBS}

Reaching 5 GHz radio fluxes $\sim 50$ mJy and $0.1 - 2.0$ keV X-ray fluxes a
few $\times 10^{-14}$ erg/cm$^2$/s, DXRBS is the faintest and largest
flat-spectrum radio sample with nearly complete ($\sim 90\%$ as of October
2000) identification. Redshift information is available for $\sim 95\%$ of the
identified sources. Starting from samples of $\sim 100,000$ sources each,
DXRBS includes only $\sim 350$ blazar candidates, which gives a measure of the
savings in terms of observing time! Moreover, our method is extremely
efficient ($\sim 90\%$ so far) at finding radio-loud quasars and BL Lacs. 

\begin{figure}
\centering
\includegraphics[width=0.6\textwidth]{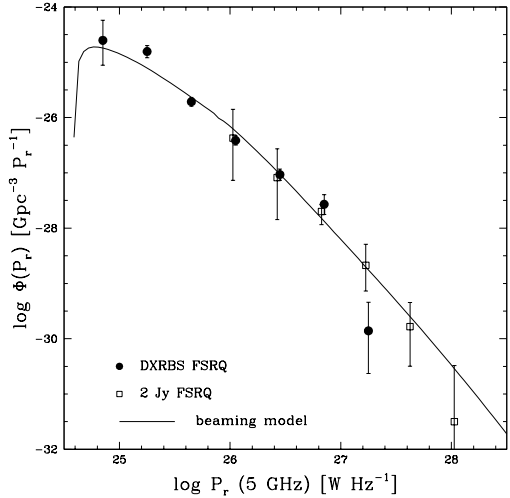}
\caption[]{The (preliminary) radio luminosity function of DXRBS FSRQ (filled
points) compared to the predictions of a beaming model based on the 2 Jy
luminosity function and evolution (solid line). The open squares represent the
2 Jy luminosity function. Error bars correspond to $1\sigma$ Poisson errors}
\label{ps1}
\end{figure}

Details on the selection technique and identification procedures can be found
in \cite{p98} and \cite{l00}, while preliminary results on the evolutionary
properties of the sample are given in \cite{p00}. Here we want to give only a
flavour of the astrophysical results that can be obtained from DXRBS in terms
of the luminosity function (LF) of FSRQ. \index{flat-spectrum radio quasars}

Figure \ref{ps1} presents the (preliminary) local radio luminosity function
(de-evolved to zero redshift using the best-fit evolution) for the DXRBS
FSRQ. We have taken into account the fact that the identification process is
not yet complete by applying the best-fit evolution derived from a complete
subsample to the whole sample. The predictions of unified schemes based on a
fit to the 2 Jy LF \cite{up95} are also shown (solid line). These basically
show what one should expect to find when reaching powers lower than those used
to constrain the luminosity function at the high end. A few interesting points
can be made: 1. the 2 Jy and DXRBS LFs are in good agreement in the region of
overlap, despite the factor $\sim 40$ difference in limiting flux; 2. DXRBS
has much better statistics: the two lowest bins of the 2 Jy LF contain only
one object each, while the number of DXRBS sources in the same bins is $\sim
20 - 30$; 3. the DXRBS LF reaches powers more than one order of magnitude
smaller than those reached by the 2 Jy LF, as expected given the much fainter
($\sim 30$) flux limit; 4. the DXRBS LF is in (amazingly!) good agreement with
the predictions of unified schemes; apparently unification of blazars and
radio galaxies seems to work even at low powers; 5. we are getting close to
the limits of the FSRQ ``Universe''; as FSRQ are thought to be the beamed
counterparts of high-power radio galaxies, their luminosity function should
end at relatively high powers. Assuming that the value inferred from the fit
to the 2 Jy LF is correct (solid line in the figure, based on the 2 Jy LF of
Fanaroff-Riley type II radio galaxies; see \cite{up95}), then DXRBS is
approaching that value.

\section{The ``Sedentary'' BL Lac Survey}

The scope of the ``sedentary'' survey is to reach deeper fluxes but only for a
subset of extreme BL Lacs, of the so-called high-energy peaked (HBL)
type. Namely, we are looking for BL Lacs with large X-ray-to-radio flux
ratios, and therefore with synchrotron peak frequency in the X-ray band (see
\cite{gmp99}, \cite{g00} for details). \index{BL Lacs}

To this aim, we cross-correlated the NVSS radio catalog \cite{c98} with the
{\it ROSAT} All Sky Survey Bright Source Catalog (RASSBSC) \cite{vo}. Optical
magnitudes were then obtained from the APM and COSMOS on-line services. This
resulted in a database of $\sim 2,000$ high Galactic latitude ($|b| >
20^{\circ}$) sources with radio, optical, and X-ray information. We then
plotted all sources on the $\alpha_{\rm ox} - \alpha_{\rm ro}$ plane. These
are the usual effective two-point spectral indices defined between the
rest-frame frequencies of 5 GHz, 5000 \AA, and 1 keV and give an overview of
the spectral energy distribution (SED) of a source. (This is particularly
useful for blazars whose SED is relatively simple, being dominated by
non-thermal processes.) It is well known that there is a region in the
$\alpha_{\rm ox} - \alpha_{\rm ro}$ plane that is almost exclusively ($\sim
90\%$) populated by HBL. We then selected all sources in this zone (delimited
by $\alpha_{\rm ro} > 0.2$ and $f_{\rm x}/f_{\rm r} > 3 \times 10^{-10}$
erg/cm$^2$/s/Jy [or $\alpha_{\rm rx} \simlt 0.56$]) and extracted a
well-defined, complete sample of 155 HBL candidates. The synchrotron peak
energy for these sources is expected to be at relatively large ($E \simgt
0.05$ keV) energies. This sample is not completely identified ($\sim 40\%$ at
the time the work was done and $\sim 70\%$ as of October 2000), but based on
the fraction of identified BL Lacs in it its HBL content is expected to be
$\sim 85\%$. Therefore, important information (number counts, evolutionary
properties via the $V_{\rm e}/V_{\rm a}$ test) can be extracted from it even
without complete optical identification. Hence the name ``sedentary'': all
this can be done while sitting in front of one's computer.

\begin{figure}
\centering
\includegraphics[width=0.6\textwidth]{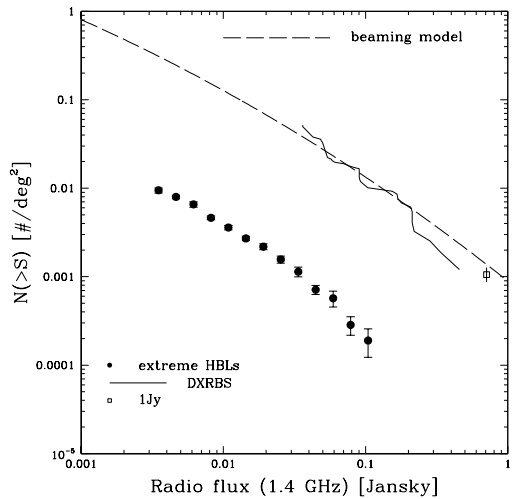}
\caption[]{The radio integral number counts of the ``sedentary'' BL Lac sample
($f_{\rm x}/f_{\rm r} > 3\times 10^{-10}$ erg/cm$^2$/s/Jy; filled
circles). The dashed line represents the expected radio counts for all types
of BL Lacs estimated from the radio luminosity function in \cite{pg95}, in
excellent agreement with the number counts of DXRBS BL Lacs (solid line). The
BL Lac surface density from the 1 Jy (open square) is also shown}
\label{ps2}
\end{figure}

Figure \ref{ps2} shows the number counts at 1.4 GHz of the ``sedentary'' BL
Lac candidates which are, as discussed above, relatively extreme, compared to
the predicted number counts for all BL Lac types (which are in good agreement
with the DXRBS counts). As explained in detail in \cite{gmp99}, the mere fact
that the shape of the ``sedentary'' counts is the same as that of the counts
for all BL Lacs (i.e., that the fraction of extreme HBL does not depend on
radio flux) {\it by itself} poses strong constraints to detailed blazar
physical models.

\section{Deeper Surveys}

The identification of radio-loud sources in X-ray and radio surveys deeper
than discussed here will pose some problems. Consider in fact that a typical
radio-loud source will have a magnitude $V \sim 24$ in a radio survey reaching
1 mJy and $V \sim 26$ at X-ray fluxes $f_{\rm x} \sim 10^{-15}$ erg/cm$^2$/s,
quite standard for Chandra/XMM observations. This is beyond the reach for
spectroscopy of 4m class telescopes even in the presence of strong, broad
lines. Furthermore, source identification at $V \sim 26$ is very time
consuming ($t_{\rm exp } \simgt 1-2$ hours) even for 8-10m class telescopes,
and becomes frustratingly difficult when dealing with an almost featureless
BL Lac. \index{statistical identification of sources} 

This means two things: 1. we will need to be very efficient in our
pre-selection of candidates, as optical identification will require large
resources; therefore, data mining will become a necessity; 2. statistical
identification of sources based on their location in multi-parameter space,
with the consequent smaller need for optical spectra (similar to the method
employed for the ``Sedentary'' survey; \S~4), will also have to become more
common. 

\section{Summary} 

The main conclusions are as follows:

\begin{itemize}

\item Blazars are very interesting astrophysical sources. By being rare and 
broad-band emitters they are also ideal for data mining studies. 

\item We have been using data mining techniques in two ways: to construct
fainter blazar samples and to find extreme blazars. In both cases data mining
is an efficient way to assemble relatively large blazar samples useful to
address (and hopefully solve) astrophysical problems. 

\item Due to the faintness of the optical counterparts, even deeper blazar
(radio-loud AGN) surveys will face daunting identification problems. Data
mining will then become a necessity and in some cases statistical
identification, based on source location in multi-parameter space, will be the
only feasible option. In the case of flat-spectrum radio quasars, however, the
good news is that we might be already approaching the limits of their
Universe. 

\end{itemize}

\leftline{\bf Acknowledgments}

The work on DXRBS reported here has been done in collaboration with, amongst
others, Hermine Landt and Eric Perlman.

\clearpage
\addcontentsline{toc}{section}{Index}
\flushbottom
\printindex

\end{document}